\documentstyle[prl,aps,twocolumn]{revtex}
\begin{document}
\draft
\wideabs{
\title{Heavy Fermion Behaviors in LiV$_{\bbox{2}}$O$_{\bbox{4}}$}
\author{D.~C. Johnston}
\address{Ames Laboratory and Department of Physics and Astronomy,
Iowa State University, Ames, Iowa 50011}
\date{October 24, 1999; Proc.\ SCES99, to be published in Physica
B}
\maketitle

\begin{abstract} Experiments of various types on the metallic
transition metal oxide compound LiV$_2$O$_4$ with the fcc
normal-spinel structure are reviewed.  The low-temperature $T
\lesssim 10$\,K data consistently indicate heavy fermion (HF)
behaviors characteristic of those of the heaviest-mass
$f$-electron HF compounds.  A crossover is observed above $\sim
50$\,K in the magnetic susceptibility,
$^7$Li spin-lattice relaxation rate and neutron magnetic
scattering function to a magnetic state variously described as an
antiferromagnetically-coupled local-moment metal or, at the
opposite extreme, as an itinerant nearly-ferromagnetic metal.
Recent theoretical investigations to understand these properties
of LiV$_2$O$_4$ are discussed.
\end{abstract}
\pacs{Keywords: Heavy fermion behaviors, LiV$_2$O$_4$} }

\section{Introduction}

The phenomenology of heavy fermion (HF) behaviors in systems
containing nearly localized $f$-electron lanthanide and actinide
atoms is by now well-developed.  These include $f$-electron
intermetallics with heavy Fermi liquid ground
states\cite{HFReviews} or non-Fermi liquid ground states
\cite{Maple1994}.  The HF
$f$-electron compounds (e.g., CeAl$_3$, UPt$_3$) have enormous
electronic specific heat coefficients $\gamma(T) \equiv C_{\rm
e}(T)/T \sim 1$\,J/mol\,K$^2$, where
$C_{\rm e}(T)$ is the temperature $T$ dependent electronic
specific heat, from which quasiparticle effective masses of
several hundred times the free electron mass have been inferred.
In such systems a high and narrow ($\sim 10$\,meV) peak occurs at
low~$T$ in the quasiparticle density of states ${\cal D}$ near the
Fermi energy $E_F$, a many-body effect
\cite{HFReviews}.  The large ${\cal D}(E_F)$ is reflected in a
large nearly
$T$-independent magnetic spin susceptibility $\chi^{\rm spin}$ and
$\gamma$ compared with the respective predictions of conventional
band structure calculations\cite{HFReviews}.  The normalized ratio
of these two quantities, the Wilson ratio $R_W$, is on the order
of unity as in conventional metals, where for spin $S = 1/2$
quasiparticles this ratio is
$R_W \equiv \pi^2 k_B^2 \chi^{\rm spin}/3 \mu_B^2 \gamma$, $k_B$
is Boltzmann's constant and $\mu_B$ is the Bohr magneton.
However, at higher $T$ the ${\cal D}(E)$ peak height decreases
strongly \cite{HFReviews}, on the scale of a low characteristic
temperature $\sim 1$--100\,K\@.  This results in a corresponding
strong decrease in $\gamma$ and $\chi^{\rm spin}$ with $T$.  HF
behaviors are not expected for $d$-electron compounds because of
the much larger spatial extent of $d$ orbitals than of $f$
orbitals and the resulting stronger hybridization with conduction
electron states.  Recently, however, HF behaviors, characteristic
of those of the heaviest mass $f$-electron HF systems, have been
found
\cite{Kondo1997} in the metallic\cite{Rogers1967} transition metal
oxide compound LiV$_2$O$_4$.

LiV$_2$O$_4$ has the face-centered-cubic normal-spinel structure
(space group
$Fd{\bar 3}m$) \cite{Reuter1960}, and is formally a $d^{1.5}$
system.  The Li atoms occupy tetrahedral holes and the V atoms
octahedral holes in a nearly cubic-close-packed oxygen
sublattice.  All V atoms and sets of any chosen clusters of V
atoms in LiV$_2$O$_4$ are respectively crystallographically
equivalent.  This crystallographic feature, together with the
non-integral oxidation state of the V atoms, result in metallic
character simply due to symmetry considerations.  Since the
structure remains undistorted down to low (4\,K) $T$
\cite{Kondo1997,Chmaissem1997}, the compound must remain metallic
upon cooling to such temperatures.  To our knowledge, the only
other spinel-structure transition metal oxide which remains
metallic to low $T$ is isostructural LiTi$_2$O$_4$, which is a
superconductor below $T_{\rm c}\leq 13.7$\,K \cite{Johnston1976}.
On the other hand, the magnetic susceptibility
$\chi(T)$ (4.2--308\,K) of LiV$_2$O$_4$ was initially found to be
the sum of a
$T$-independent term and a Curie-Weiss term $C/(T - \theta)$ due
to local V magnetic moments \cite{Kessler1971}.  The Curie
constant $C$ is consistent with a V$^{+4}$ spin $S = 1/2$ with
$g$-factor 2.23, suggesting that one of the 1.5
$d$-electrons/V atom is (nearly) localized on the atom.  The
negative Weiss temperature
$\theta = -63$\,K indicates antiferromagnetic (AF) V spin
interactions, and the anomalously large $g$-factor suggests
ferromagnetic interactions between the local moments and the
conduction electrons \cite{Johnston1995}.  No evidence for
magnetic ordering above 4.2\,K was found.  The local moment
behavior strongly contrasts with the relatively $T$-independent
Pauli paramagnetism of LiTi$_2$O$_4$ \cite{Johnston1976}.

The metallic character of LiV$_2$O$_4$ together with the above
evidence for local moment magnetism motivated the experiments
which led to the discovery\cite{Kondo1997} of HF behaviors in this
compound.  In particular, we reasoned that due to the geometric
frustration for AF ordering inherent in the V sublattice of the
structure, perhaps long-range magnetic ordering would be
completely suppressed.  In this case, a potentially interesting
and unusual ground state would be expected.  In the following
section, the experimental results which have established the HF
character of LiV$_2$O$_4$ at low $T$, and a crossover to a
different magnetic state above $\sim 50$\,K, will be reviewed.
Then recent theoretical studies to understand these  behaviors
will be briefly discussed followed by concluding remarks.

\section{Review of Experimental Results}

When screening materials for HF behaviors, the initial experiments
are usually specific heat at constant pressure $C_{\rm p}(T)$ and
$\chi(T)$ measurements, and also resistivity $\rho(T)$
measurements.  In the case of LiV$_2$O$_4$, single crystals were
not initially available and $\rho(T)$ measurements on
polycrystalline oxides are notoriously unreliable quantitatively.
Before beginning detailed
$\chi(T)$ and $C_{\rm p}(T)$ measurements, we needed to confirm
that the Curie-Weiss
$\chi(T)$ behavior cited above, which was observed from 300\,K
down to 4\,K, was intrinsic to the material.  After an extensive
synthesis effort to reduce the concentration of paramagnetic
defects, we found that it was not. The $\chi(T)$ does show a
Curie-Weiss-like behavior from 400\,K down to $\sim 50$\,K, with
$C$ and
$\theta$ parameters similar to those cited above, but the intrinsic
$\chi(T)$ was found to level off below 30\,K, showing a weak broad
maximum at $\approx 16$\,K \cite{Kondo1997,Kondo1999}.  This
intrinsic behavior of high magnetic-purity samples was found
independently by two other groups \cite{Ueda1997,Onoda1997}.
Zero-field-cooled and field-cooled magnetization $M$ measurements
in low ($\sim 100$\,G) magnetic fields $H$ showed no hysteresis
above 2\,K, indicating that the broad peak in
$\chi(T)$ is not due to static short-range spin-glass ordering
\cite{Kondo1997,Kondo1999}.  Positive muon spin
rotation/relaxation $\mu$SR measurements on one of the highest
purity polycrystalline samples showed no evidence for static
magnetic ordering above 0.02\,K
\cite{Kondo1997}.  A sample with a significantly higher magnetic
impurity level showed static spin-glass-type ordering at about
0.8\,K\@.  $\rho(T)$ measurements on a polycrystalline pellet down
to 0.01\,K showed no evidence for superconductivity
\cite{Kondo1997}.

$C_{\rm p}(T)$ measurements have been carried out on a variety of
polycrystalline LiV$_2$O$_4$ samples
\cite{Kondo1997,Johnston1999}.  The highest magnetic purity
samples, which have magnetic impurity concentrations corresponding
to less than 0.1\,mol\% of $S = 1/2$ impurities, yielded
$\gamma(1$\,K)$\approx 0.42$--0.43\,J/mol\,K$^2$, which is the
largest value observed for any metallic nonmagnetic $d$-metal
compound.  The effective mass of the quasiparticles, assuming a
single spherical band model and a carrier concentration of 1.5
electrons/V atom, is about 180 free electron masses.  The Wilson
ratio is $R_{\rm W} \approx 1.7$ assuming quasiparticles with $S =
1/2$ and $g\approx 2$.  These results are consistent with a heavy
Fermi liquid picture at low $T$.  The $C_{\rm e}(T)$ was
determined by subtracting the lattice specific heat contribution
(that determined for LiTi$_2$O$_4$) from $C_{\rm p}(T)$.
$\gamma(T)$ decreases rapidly with
$T$ up to
$\approx 30$\,K, then decreases much more slowly up to 108\,K\@.
$C_{\rm p}(T)$  measurements on single crystals confirmed the
above low-$T$ behavior
\cite{Takagi1999}.  Polycrystalline samples with significantly
larger magnetic impurity concentrations showed an increase in
$C_{\rm p}(T)$ with decreasing $T$ below 4\,K
\cite{Kondo1997,Johnston1999} instead of leveling off with
decreasing $T$ as in the purest samples.

The linear thermal expansion coefficient $\alpha(T)$ of
LiV$_2$O$_4$ was first  measured using high resolution neutron
diffraction from room temperature to 4\,K
\cite{Chmaissem1997}.  These measurements showed normal behavior
above 60\,K, but below this temperature $\alpha(T)$ became nearly
independent of $T$, and then below 20\,K
$\alpha(T)$ strongly increased again with decreasing $T$ down to
4\,K\@.  The low-$T$ results are qualitatively as expected on the
basis of the known $f$-electron HF phenomenology.  High resolution
$\alpha(T)$ measurments were subsequently carried out using
differential capacitance dilatometry \cite{Johnston1999} which
confirmed the neutron diffraction results.  From the $\alpha(T)$
and $C_{\rm p}(T)$ data on the same sample, the Gr\"uneisen
parameter $\Gamma(T)\sim
\alpha(T)/C_{\rm p}(T)$ was determined from 4\,K to 108\,K
\cite{Johnston1999}.  At the lowest $T$, the lattice contribution
to
$\Gamma(T)$ becomes negligible compared to the electronic
contribution $\Gamma_{\rm e}(T)$.  The $\Gamma(T)$ increases
rapidly with decreasing $T$ below $\approx 20$\,K, extrapolating
to $\Gamma_{\rm e}(0)\approx 11.4$, which is intermediate between
those of conventional nonmagnetic metals and $f$-electron HF
compounds.  Thus these measurements have identified two additional
thermodynamic quantities in LiV$_2$O$_4$, the electronic
contribution to the thermal expansion coefficient and the
electronic Gr\"uneisen parameter, which have $T$ dependences
similar to those in heavy-mass
$f$-electron HF compounds.

The spin dynamics in LiV$_2$O$_4$ has been investigated using
$^7$Li nuclear magnetic resonance (NMR) measurements
\cite{Kondo1997,Onoda1997,Fujiwara1997,Mahajan1998}.  The
spin-lattice relaxation rate $1/T_1(T)$ shows a broad maximum at
30--50\,K and becomes nearly independent of $T$ from 500\,K to
800\,K\@.  These high-$T$ data above 50\,K were interpreted by one
group as indicating that at high $T$, LiV$_2$O$_4$ is a nearly
ferromagnetic metal \cite{Fujiwara1997}, whereas another group
including the present  author interpreted the data as indicating
that this compound is at the opposite end of the
localized-itinerant magnetism spectrum, namely a metal containing
nearly localized magnetic moments \cite{Mahajan1998}.  A study of
$1/T_1(T)$ from 1.5\,K to 4.2\,K for five samples with different
magnetic impurity concentrations showed a robust proportional
dependence of $1/T_1$ on
$T$ \cite{Kondo1997,Mahajan1998}.  The slope for the purest
samples, $1/T_1 T \approx 2.25$\,sec$^{-1}$K$^{-1}$, is about 4100
times greater than in LiTi$_2$O$_4$, consistent with a HF
interpretation.  The Korringa ratio, $R_{\rm K} = K^2 T_1 T/S_{\rm
Li}$ where $K$ is the Knight shift and $S_{\rm Li}$ is a
constant, was found in this $T$ range to be about 0.6, i.e.\ on
the order of unity as expected for a Fermi liquid.

The spin dynamics has also been investigated using quasielastic
neutron scattering measurements on a polycrystalline sample of
LiV$_2$O$_4$
\cite{Krimmel1999}.  The quasielastic linewidth $\Gamma$ at 15\,K
was found to be independent of scattering wavevector, with a
finite magnitude for $T\to 0$.  With increasing $T$, $\Gamma$
first showed a minimum at about 10\,K, then increased
approximately as $\sqrt{T}$.  These are characteristic features of
conventional
$f$-electron HF compounds.  Above $\sim 40$\,K, a dramatic
crossover in the dynamical structure factor $S(Q,\omega)$ was
found, indicative of a different type of metallic state  which the
authors interpreted as a possible nearly ferromagnetic metal, in
agreement with the conclusion of Ref.~\cite{Fujiwara1997}.

Single-crystal $\rho(T)$ and Hall effect measurements have been
carried out to low (0.3\,K) temperatures \cite{Takagi1999}.  The
$\rho(T)$ data show a $T^2$ dependence at low $T$, with a large
coefficient, as expected for a HF Fermi liquid.

High-field $M(H)$ measurements on high-purity polycrystalline
LiV$_2$O$_4$ were performed in static
$H$ up to 20\,T and in pulsed $H$ up to 60\,T \cite{Lacerda}.
From 0.6\,K to 50\,K,
$M\propto H$ up to about 17\,T, with a slope vs $T$ in agreement
with the $\chi(T)$ data.  Upon further increasing $H$ at 0.6\,K,
$M(H)$ shows positive curvature and then a sharp inflection point
at $H_{\rm c} = 42$\,T suggestive of a phase transition.  At even
higher fields, $H\gtrsim 50$\,T, $M(H)$ exhibits a plateau with
$M\approx 0.6\,\mu_{\rm B}$/V atom.  This $M$ is less than the
$d^1$ local-moment value of $\approx 1$\,$\mu_{\rm B}$/V atom.

Characteristic temperatures $T^\ast$ associated with the various
HF behaviors in LiV$_2$O$_4$ can be obtained from the respective
measurements discussed above.  The $\rho(T)$ data for single
crystals \cite{Rogers1967} exhibit a smooth but pronounced
downturn upon cooling below $T^\ast \approx 30$\,K\@.  The smooth
maximum in $\chi(T)$ occurs at $T^\ast\approx 16$\,K
\cite{Kondo1997,Kondo1999}.  An analysis of the $C_{\rm e}(T)$
data in terms of the single-ion Kondo model leads to $T^\ast =
26.4$\,K
\cite{Kondo1997,Johnston1999}.  Similarly, the entropy relation
$\gamma(T=1\,{\rm K})T^\ast = 2R\ln 2$ ($R$ is the molar gas
constant and the prefactor of 2 comes from 2 V atoms per formula
unit) yields
$T^\ast = 27$\,K\@.  Strong upturns with decreasing $T$ in
$\gamma(T)$, $\alpha(T)/T$ and in the electronic Gr\"uneisen
parameter occur below $T^\ast = 20$--30\,K
\cite{Kondo1997,Johnston1999}.  Equating the magnetic field energy
$\mu_{\rm B}H_{\rm c}$ at the inflection field $H_{\rm c}$ in the
high-field $M(H)$ measurements at 0.6\,K with $k_{\rm B}T^\ast$
gives $T^\ast = 28$\,K \cite{Lacerda}.  In contrast to the above
values of $T^\ast$, from the quasielastic neutron scattering
measurements a much lower Kondo-lattice temperature $T^*\approx
5$--10\,K has been inferred
\cite{Krimmel1999}.

\section{Recent Theory}

In an ionic picture, the five-fold 3$d$ orbital degeneracy of the
isolated V atom is lifted in LiV$_2$O$_4$ due to the nearly cubic
crystalline electric field (CEF) seen by the V atoms, yielding a
ground triply-degenerate
$t_{2g}$ manifold and an excited doubly-degenerate $e_g$ manifold
separated by
$\sim 2$\,eV\@.  The slight trigonal distortion of the cubic CEF
in the actual structure slightly splits the ground $t_{2g}$
triplet into an $E_g$ doublet and an $A_{1g}$ singlet.

Four LDA band structure calculations have been reported for
LiV$_2$O$_4$
\cite{Eyert1999,Matsuno1999,Anisimov1999,Singh1999}.  The
calculations consistently show a metallic ground state. The
ionic picture just described is strongly reflected in the band
structure.  $E_{\rm F}$ lies within bands derived from the
$t_{\rm 2g}$ orbitals; these bands are separated from those
derived from the $e_g$ orbitals by an energy of $\approx
1$\,eV\@.  The O 2$s$ and 2$p$ bands are filled, lying below the
$t_{2g}$ bands, whereas the Li 2$s$ bands far above $E_{\rm F}$
are empty, consistent with a $-2$ oxidation state of O and a $+1$
oxidation state of Li in the ionic picture.  The slight splitting
of the $E_g$- and
$A_{1g}$-derived bands is not readily discernable in the band
calculations.  From the band calculations
\cite{Matsuno1999,Singh1999}, the ${\cal D}(E_{\rm F}) \approx
7.2$\,states/eV\,formula\,unit is about a factor of 25 smaller
that the experimental value inferred from $C_{\rm p}(T)$ at low
$T$, indicating that electron correlation effects are important in
LiV$_2$O$_4$.

Electron correlation effects have been discussed using several
ideas and approaches.  For example, Eyert {\it et
al.}\cite{Eyert1999} favor a model in which the geometric
frustration for AF ordering of local V moments leads to a large
enhancement of spin fluctuations at low $T$, in turn resulting in
heavy-mass quasiparticles.  This is a very different picture than
for conventional $f$-electron HF compounds, in which hybridization
of local-moment orbitals with those of the conduction electrons
lead to mass enhancement via a many-body effect.  Anisimov {\it et
al.}\cite{Anisimov1999} have investigated a
model\cite{Kondo1999,Johnston1999,Mahajan1998} of the latter type
in which one of the 1.5 $d$-electrons/V atom is localized in the
$A_{1g}$ orbital due to electron correlation effects, with the
remaining 0.5 $d$-electron/V atom being itinerant in broad
$E_g$-derived bands, leading to the observed metallic
character.  In such a model, ferromagnetic double exchange between
the local moments would be important and the outstanding question
would then be why LiV$_2$O$_4$ is not a metallic ferromagnet at
low $T$
\cite{Anisimov1999,Varma1999}.  Anisimov {\it et al.}\ found that
within the LDA$+$U approach one electron is indeed localized in an
$A_{1g}$ orbital on each V atom.  The effects of double exchange
were found to be largely canceled by an off-site AF Kondo-like
exchange.  They solved an effective Anderson impurity model, which
when combined with the ``exhaustion phenomenon'' of Nosi\`eres,
led to a Kondo lattice model with a coherence temperature $T^\ast$
comparable to the experimental values.

Much progress has been made theoretically to understand the
properties of LiV$_2$O$_4$.  Quantitative calculations of the $T$
dependences of physical observables and comparisons with
experimental results will ultimately determine which of the
current views of electron correlation effects in  LiV$_2$O$_4$ are
most appropriate.

\section{Concluding Remarks}

In this review we have referred always to heavy Fermi liquid {\it
behaviors} of LiV$_2$O$_4$ at low $T$, i.e.\ within the $T$ ranges
of the various measurements, and not to a heavy Fermi liquid {\it
ground state}.  It is of course possible that the intrinsic ground
state is not a Fermi liquid.

There is no theoretical consensus yet about the mechanism(s)
for the low-$T$ HF behaviors of LiV$_2$O$_4$.  In addition, a
theoretical description of the crossover to the high-$T$ behaviors
starting at 30--50\,K remains to be given.  The origin of the
high-$T$ behaviors themselves remains to be clarified, in
particular whether this compound should be more appropriately
thought of at high temperatures as being an itinerant nearly
ferromagnetic metal or a metal containing nearly localized
magnetic moments.  These states lie at
opposite ends of the localized vs itinerant magnetism spectrum.
Perhaps the correct view is somewhere in between.

Several features of LiV$_2$O$_4$, some of which were discussed
above, may be involved in the mechanism for the HF behaviors at
low $T$.  First, in a local moment model, geometric frustration
for AF ordering is present in the V atom substructure which would
be expected to suppress static magnetic ordering.  Second, the
same  three $t_{2g}$ orbitals give rise to the combined metallic
and local moment behaviors at high $T$ and the HF behaviors at low
$T$.  This is in strong contrast to
$f$-electron HF compounds where the nearly-localized $f$ orbitals
are clearly distinguished from the strongly delocalized $s$, $p$
and/or $d$ conduction band states.  Third, due to the high
symmetry of the V sites, two of the three partially filled
$t_{2g}$ orbitals are strictly degenerate from symmetry
considerations and are nearly degenerate with the third, so
dynamical orbital ordering and the charge degrees of freedom may
be expected to be important.  Finally, because the spin
interactions between neighboring V atoms strongly depend on which
of the respective $d$-orbitals are occupied, the spin degrees of
freedom would be expected to be strongly coupled with dynamical
orbital ordering.  Thus the observed low-$T$ HF behaviors may
involve coupled dynamical charge, orbital and spin correlations,
which may all be affected by the geometric frustration for AF
ordering.  This is a complex scenario difficult to treat
theoretically.

The physical properties of LiV$_2$O$_4$ are very different from
those of LiTi$_2$O$_4$, an isostructural superconductor with a
conventional $d$-band ${\cal D}(E_{\rm F})$.  The LDA band
structures of both compounds are very similar.  A successful
theoretical framework to explain the low- and high-$T$ physical
properties of the $d^{1.5}$ compound LiV$_2$O$_4$ must also be
able to explain why these properties are so divergent from those
of the $d^{0.5}$ compound LiTi$_2$O$_4$.  A beginning has been
made by Varma\cite{Varma1999} and implicitly as discussed in the
previous section.  A successful framework will lead to deeper
understanding of metallic transition metal oxides generally and
perhaps also of the $f$-electron class of heavy fermion compounds.

\acknowledgments

The author is very grateful to his collaborators in much of the
work cited here and to many colleagues for stimulating
discussions.  Ames Laboratory is  operated for the U.S. Department
of Energy by Iowa State University under Contract No.\
W-7405-Eng-82.  This work was supported by the Director for Energy
Research, Office of Basic Energy Sciences.


\begin{references}

\bibitem{HFReviews}G. R. Stewart, Rev.\ Mod.\ Phys.\ {\bf 56}, 755
(1984); A.~C.~Hewson, {\it The Kondo Problem to Heavy Fermions}
(Cambridge University Press, Cambridge, 1993).

\bibitem{Maple1994}M. B. Maple {\it et al.}, J. Low Temp.\ Phys.\
{\bf 95}, 225 (1994).

\bibitem{Kondo1997}S. Kondo, D. C. Johnston, C. A. Swenson, F.
Borsa, A. V. Mahajan, L. L. Miller, T. Gu, A. I. Goldman, M. B.
Maple, D. A. Gajewski, E. J. Freeman, N. R. Dilley, R. P. Dickey,
J. Merrin, K. Kojima, G. M. Luke, Y. J. Uemura, O. Chmaissem, and
J. D. Jorgensen, Phys.\ Rev.\ Lett.\ {\bf 78}, 3729 (1997).

\bibitem{Rogers1967}D. B. Rogers, J. L. Gillson, and T. E. Gier,
Solid State Commun.\ {\bf 5}, 263 (1967).

\bibitem{Reuter1960}B. Reuter and J. Jaskowsky, Angew.\ Chem.\
{\bf 72}, 209 (1960); Ber.\ Bunsen-Ges.\ Phys.\ Chem.\ {\bf 70},
189 (1966).

\bibitem{Chmaissem1997}O. Chmaissem, J. D. Jorgensen, S. Kondo,
and D. C.
\mbox{Johnston}, Phys.\ Rev.\ Lett.\ {\bf 79}, 4866 (1997).

\bibitem{Johnston1976}D. C. Johnston, H. Prakash, W. H.
Zachariasen, and R. Viswanathan, Mater.\ Res.\ Bull.\ {\bf 8}, 777
(1973); D. C. Johnston, J. Low Temp.\ Phys.\ {\bf 25}, 145 (1976).

\bibitem{Kessler1971}H. Kessler and M. J. Sienko, J. Chem.\ Phys.\
{\bf 55}, 5414 (1971); see also Y. Nakajima {\it et al.}, Physica
(Amsterdam) {\bf 185C-189C}, (1991) and
Ref.~\protect\cite{Johnston1995}.

\bibitem{Johnston1995}D. C. Johnston, T. Ami, F. Borsa, M. K.
Crawford, J. A. Fernandez-Baca, K. H. Kim, R. L. Harlow, A. V.
Mahajan, L. L. Miller, M. A. Subramanian, D. R. Torgeson, and Z.
R. Wang, Springer Ser.\  Solid-State Sci.\ {\bf 119}, 241 (1995).

\bibitem{Kondo1999}S. Kondo, D. C. Johnston, and L. L. Miller,
Phys.\ Rev.\ B {\bf 59}, 2609 (1999).

\bibitem{Ueda1997}Y. Ueda, N. Fujiwara, and H. Yasuoka, J. Phys.\
Soc.\ Jpn.\ {\bf 66}, 778 (1997).

\bibitem{Onoda1997}M. Onoda, H. Imai, Y. Amako, and H. Nagasawa,
Phys.\ Rev.\ B {\bf 56}, 3760 (1997).

\bibitem{Johnston1999}D. C. Johnston, C. A. Swenson, and S. Kondo,
Phys.\ Rev.\ B {\bf 59}, 2627 (1999).

\bibitem{Takagi1999}H. Takagi {\it et al.}, unpublished.

\bibitem{Fujiwara1997}N. Fujiwara, Y. Ueda, and H. Yasuoka,
Physica B {\bf 237-238}, 59 (1997); N. Fujiwara, H. Yasuoka, and
Y. Ueda, Phys. Rev. B {\bf 57}, 3539 (1998).

\bibitem{Mahajan1998}A. V. Mahajan, R. Sala, E. Lee, F. Borsa, S.
Kondo, and D. C. Johnston, Phys.\ Rev.\ B {\bf 57}, 8890 (1998).

\bibitem{Krimmel1999}A. Krimmel, A. Loidl, M. Klemm, S. Horn, and
H. Schober, Phys.\ Rev.\ Lett.\ {\bf 82}, 2919 (1999).

\bibitem{Lacerda}D. C. Johnston, A. Lacerda, S. Kondo, {\it et
al.}, unpublished.

\bibitem{Eyert1999}V. Eyert, K.-H. H\"ock, S. Horn, A. Loidl, and
P. S. Riseborough, Europhys.\ Lett.\ {\bf 46}, 762 (1999).

\bibitem{Matsuno1999}J. Matsuno, A. Fujimori, and L. F. Mattheiss,
Phys.\ Rev.\ B {\bf 60}, 1607 (1999); J. Matsuno, K. Kobayashi, A.
Fujimori, L. F. Mattheiss, and Y. Ueda, Physica B (to be
published).

\bibitem{Anisimov1999}V. I. Anisimov, M. A. Korotin, M. Z\"olfl,
T. Pruschke, K. Le Hur, and T. M. Rice, Phys.\ Rev.\ Lett.\ {\bf
83}, 364 (1999).

\bibitem{Singh1999}D. J. Singh, P. Blaha, K. Schwarz, and I. I.
Mazin, cond-mat/9907206 v2 (unpublished).

\bibitem{Varma1999}C. M. Varma, Phys.\ Rev.\ B {\bf 60}, R6973
(1999).

\end{references}
\end{document}